\documentclass{ws-rv9x6}
\usepackage{subfigure}
\usepackage[latin1]{inputenc}

\usepackage{color,soul}
\setstcolor{red}
\usepackage{ws-rv-van}             % numbered citation/references (default)
\makeindex
\usepackage{amsmath,amssymb}

\newcommand{\be}{\begin{equation}}
\newcommand{\ee}{\end{equation}}

\begin{document}

\chapter[Arrival statistics and exploration properties of mortal walkers]{Arrival statistics and exploration properties of mortal walkers}
\label{chLindenberg}

\author{Santos B. Yuste\footnote{santos@unex.es}}
%\author{Primer Author\footnote{Author footnote\oint.}}
\address{Departamento de F\'{\i}sica, Universidad de Extremadura, E-06071 Badajoz, Spain}
%\address{Primer author's address\footnote{Affiliation footnote.}}
\author{E. Abad\footnote{eabad@unex.es}}
%\author[Nombre autores para Running Head: F. Author and S. Author]{Second Author}
\address{Departamento de F\'{\i}sica Aplicada, Centro Universitario de Mérida,
Universidad de Extremadura, E-06800 Mérida, Spain}
%\address{Second author's address}++++
\author[S. B. Yuste, E. Abad and K. Lindenberg]{Katja Lindenberg\footnote{klindenberg@ucsd.edu}}
\address{Department of Chemistry and Biochemistry and BioCircuits Institute, University of
California San Diego, La Jolla, California 92093-0340, USA}

\begin{abstract}
We study some of the salient features of the arrival statistics and exploration properties of mortal random walkers, that is, walkers that may die as they move, or as they wait to move. Such evanescence or death events have profound consequences for quantities such as  the number of distinct sites visited which are relevant for the computation of encounter-controlled rates in chemical kinetics. We exploit the observation that well-known methods developed decades ago for immortal walkers are widely applicable to mortal walkers. The particular cases of exponential and power-law evanescence are considered in detail. Finally, we discuss the relevance of our results to the target problem with mortal traps and a particular application thereof, namely, the defect diffusion model. Evanescence of defects is postulated as a possible complementary contribution or perhaps even an alternative to anomalous diffusion to explain observed stretched exponential relaxation behavior.
\end{abstract}
%\markright{Customized Running Head for Odd Page} % default is chapter title.
\body

\section{Introduction}

Random walks play a central role in statistical physics as elementary models describing stochastic transport in a large variety of experimental systems. \cite{BookWeiss} While there is a large body of literature dealing with the first-passage and exploration behavior of walkers that do not undergo reactions or transformations, considerably less attention has been devoted to the corresponding properties of reactive walkers. In particular, work involving exploration properties of so-called mortal or evanescent walkers, that is, walkers that die in the course of their motion, is surprisingly scarce despite the ubiquitous occurrence of such death processes in many physical, chemical, and biological scenarios.

 \emph{Ad-hoc} models for particular experimental situations have previously been implemented, e.g., to deal with photon scattering and absorption in a tissue \cite{Bonner}. And yet, with very few exceptions (see e.g. Refs. \citen{Bonner, Lohmar} or subsection 3.2.4 in Ref. \refcite{Hughes}). the observation that the effect of elementary death processes on the exploration properties of random walkers can be investigated using arguments similar to those used for immortal walkers seems until recently to have gone largely unnoticed. \cite{YALPRL13} The aim of the present work is to introduce in a pedagogical way an overarching theoretical framework allowing one to obtain the most interesting properties of mortal walkers. In view of the completely new physics introduced by death processes in the transport behavior and its potential relevance for experiments such as for instance the defect relaxation system discussed in Sec. \ref{Sec:tp}, this endeavor seems very timely. 
 
 In what follows we present in a more general framework some recent results first given in Ref. \refcite{YALPRL13} in a succint fashion. More specifically, we shall focus on properties characterizing the arrival statistics of random walkers in discrete time, such as the average number of distinct sites visited after a certain number of steps on perfect lattices of all dimensions and other related quantities. In the continuum time limit one can establish a relationship between these quantities and the number of distinct sites visited up to a given time (rather than step number), a key quantity for the computation of reaction rates in chemical kinetics.

Mortal random walkers have a probability of dying as they walk.  This means that their density $\rho$ is not a constant as it is in an ordinary random walk but is instead a decaying function of time, $\rho(t)$, or of step number, $\rho(n)$.  One can think of this decay as arising, for instance, from a chemical process that removes the walkers or transforms them into an inert species. This might be the result of a spontaneous random process, e.g., a radioactive decay, or of an interaction with other particles, as is  the case, for example, in scavenging reactions. We will later specify particular decay functions $\rho(n)$ or $\rho(t)$.

The present chapter is organized as follows. In Sec. \ref{Sec:GenFun} we show how to obtain some properties related to the arrival statistics of mortal walkers at given lattice sites in the framework of the generating function formalism originally developed by Montroll and Weiss \cite{Montroll1965} and extended in subsequent work (see for example chapter 6 in Ref. \refcite{Hughes} and references therein). Next, in Sec. \ref{Sec:Res}, we discuss the behavior of some of these quantities for particular forms of the decay law of the walker density, namely, exponential decay and power-law decay. We conclude that the behavior in the case of mortal walkers is very different from that observed for immortal walkers. For instance, as time goes to infinity the number of distinct sites visited tends to infinity in the latter case, while in the former case it goes to a finite value provided that the evanescence is sufficiently fast. In Sec. \ref{Sec:tp} we discuss the implications of these results for the target problem with mortal traps, in particular, the possibility that the target survives forever if the traps die sufficiently rapidly. On the basis of our results, we suggest that the stretched exponential behavior observed in some relaxational systems driven by defect diffusion can be ascribed to evanescence without the need to invoke anomalous diffusion. Finally, in Sec. \ref{Sec:Conc} we briefly summarize our main conclusions and discuss possible extensions of our work.
The appendices are devoted to some technicalities related to the correspondence between the connection of the master equation for mortal walkers to the continuum reaction-diffusion equation with a linear reaction term, as well as to the on-lattice kinetics of mortal walkers and continuum space exploration.

\section{Generating function formalism for mortal walkers}
\label{Sec:GenFun}

\subsection{Master equation for mortal walkers}

Our starting point is a mortal walker which performs a P\'olya walk, i.e., a symmetric nearest-neighbor walk on a perfect lattice of dimension $d$.  We introduce the probability $P^*_{m,n-m}(s|s')$ of finding a mortal walker at lattice site $s$ after having taken $n-m$ steps if the walk started at site $s'$ at step $m$.  An asterisk denotes quantities for \emph{mortal} walkers and the same quantities without asterisks are for immortal walkers. The probability $P^*$ for a mortal walker is related to its counterpart $P$ for an immortal walker as follows:
\be\label{P*Pgb}
P_{m,n-m}^*(s|s')=\frac{\rho(n)}{\rho(m)} \,P_{m,n-m}(s|s') = \frac{\rho(n)}{\rho(m)} \,P_{n-m}(s|s'),
\ee
where the second equality follows from the fact that in the absence of the death process the probability does not depend on when the walk started but only on the number of steps taken by the walker.  When $m=0$ we can  more simply write
\be
\label{P*P0}
P_{n}^*(s|s_0)=\frac{\rho(n)}{\rho(0)}  \,P_{n}(s|s_0),
\ee
where we have simplified the notation,  $P_{0,n}^*(s|s_0)\equiv P_{n}^*(s|s_0)$. The walker density ratio
$\rho(n)/\rho(0)$ can be interpreted as the survival probability of the walker up to step $n$, or as the fraction of surviving walkers if one considers a statistical ensemble of walkers all with the same initial condition. In what follows, we set $\rho(0)=1$, and consequently $P_{n}^*(s|s_0)=\rho(n)  \,P_{n}(s|s_0)$. For simplicity, we also restrict ourselves to the deterministic initial condition
\be
P_0^*(s|s_0)=\delta_{ss_0}.
\ee
From the well-known master equation for the sojourn probabilities for an immortal walker
and the definition \eqref{P*P0} one gets the corresponding master equation for mortal walkers. For instance,
in $d=1$ one has
\be
P_{n+1}(s|s_0)=\frac{1}{2}\left[P_n(s-1|s_0)+P_n(s+1|s_0)\right]
\ee
and, as a result of Eq. \eqref{P*P0},
\be
\label{evanescentME}
P_{n+1}^*(s|s_0)=\frac{1}{2}\frac{\rho(n+1)}{\rho(n)}\left[P_n^*(s-1|s_0)+P_n^*(s+1|s_0)\right].
\ee
This equation simply states that the probability that the particle is found at site $s$ after $n+1$ time steps is equal to the probability that it was either on the left neighbor site $s-1$ or the right neighbor site $s+1$ at the previous time step, then chose to jump to $s$ and did not vanish while doing so (the probability for the particle to survive up to the $n+1$-st step given that it had survived up to the $n$th step is simply $\rho(n+1)/\rho(n)$).
As shown in Appendix \ref{appRDE}, taking the diffusive limit of \eqref{evanescentME} yields a continuum reaction-diffusion equation with a linear reaction term. While the reaction-diffusion equation is amenable to exact solution, we shall not pursue this route in what follows, as we are interested in retaining the discrete spatial structure of the lattice for the sake of studying some aspects of the arrival statistics of a mortal walker at particular lattice sites.

\subsection{First-passage probabilities and statistics of site occupation}

Next, we introduce the probability $F^*_n(s|s_0)$ that the evanescent walker arrives at site $s$ \emph{for the first time} at step $n$ if she started at site $s_0$ at step $n=0$. The first arrival probabilities in the presence and absence of evanescence are also simply related to one another,
\be\label{F*F}
F_n^*(s|s_0)=\rho(n) \,F_n(s|s_0),
\ee
and the initial condition is as for immortal walkers,
\be
F_0^*(s|s_0)=0.
\ee
The probabilities $F^*$ and $P^*$ are related to one another in the same way as in a non-evanescent walk,
\be
\label{P*FP}
P_n^*(s|s_0)= \sum_{j=1}^n F_j^*(s|s_0) P_{j,n-j}^*(s|s),\quad n\ge 1.
\ee
In other words, it still holds that the probability that the evanescent walker is at site $s$ after $n$ steps (without having evanesced) can be decomposed into the $n$ mutually exclusive events that the walker first arrived at site $s$ after $j$ steps (without having evanesced) and returned to site $s$ in $n-j$ steps (without having evanesced).

Later we will also find it convenient to use the generating functions for these two probabilities,
\begin{align}
P^*(s|s_0;\xi)&=\sum_{n=0}^\infty P_n^*(s|s_0)\xi^n=\sum_{n=0}^\infty P_n(s|s_0)\rho(n)\xi^n,\\
F^*(s|s_0;\xi)&=\sum_{n=0}^\infty F_n^*(s|s_0)\xi^n=\sum_{n=0}^\infty F_n(s|s_0)\rho(n)\xi^n.
\end{align}

\subsubsection{Number of distinct sites visited by a mortal walker}

We now introduce two additional quantities, namely, the number of distinct sites visited in an $n$-step evanescent walk, $S_n^*$, and the number of new sites $\Delta_n$ (sites never visited before) visited in the $n$th step of an evanescent walk.  We follow the convention that the start of the walk counts as a visit to the starting site $s_0$, i.e.,  $\Delta_0^*=1$. The following relations are fairly obvious. The first relates walks with and without evanescence,
\be
\label{D*D}
\langle \Delta^*_n \rangle  =\rho(n)\, \langle \Delta_n \rangle,
\ee
where the brackets denote averages over realizations of the random walk.
Clearly, $\langle \Delta_n \rangle$ is a number between zero and one that can be interpreted as the probability to visit a new site at the $n$-th step. We next relate evanescent walk quantities to one another as follows,
\be
\label{SDelta}
S^*_n=\sum_{j=0}^n \Delta^*_j   \quad \leftrightarrow \quad \langle S^*_n \rangle=\sum_{j=0}^n \langle\Delta^*_j \rangle,
\ee
\be
\label{SDelta2}
 \langle\Delta^*_n \rangle= \langle S^*_n \rangle - \langle S^*_{n-1} \rangle.
\ee
Let us further introduce the two generating functions associated with these two quantities, namely,
\be
S^*(\xi)= \sum_{n=0}^\infty   \langle  S^*_n \rangle  \xi^n
\ee
and
\be
\Delta^*(\xi)=\sum_{n=0}^\infty \langle \Delta^*_n \rangle  \xi^n.
\ee

It is useful to establish relations between some of the generating functions. First, multiplying the left equation in Eq.~(\ref{SDelta}) by $\xi^n$, summing over $n$, and changing the order of summations leads to
\be
\label{SD2}
S^*(\xi)= \frac{\Delta^*(\xi)}{1-\xi}.
\ee
Next, note that
\be
\label{DPrF}
\langle\Delta^*_j \rangle=\text{Pr}(\Delta^*_j=1)=\sum_{s\neq s_0} F^*_j(s|s_0).
\ee
Again multiplying by $\xi^n$, summing over $n$, and reversing the order of summation leads to
\be
\label{D*F}
\Delta^*(\xi) = 1+\sum_{s\neq s_0}   F^*(s|s_0;\xi).
\ee

We now follow with the important observation that  Eq.~(\ref{SDelta}) with (\ref{D*D}) allows one to calculate the average number of distinct sites visited by a mortal walker up to step $n$ in terms of non-evanescent walk properties,
\be
\langle S_n^*\rangle = \sum_{j=0}^n \rho(j)\langle \Delta_j \rangle.
\label{Scalc}
\ee
On the other hand, it is well-known from the theory for immortal walkers that the $\langle \Delta_j \rangle$'s fulfil the recursion relation (see e.g. Ref. \refcite{Hughes}, pp. 324-25)
\be
\label{Dnr}
\langle \Delta_j\rangle =1 -\sum_{k=1}^j P_k(0) \langle \Delta_{j-k}\rangle
\;,
\quad
\text{with}
\quad
 j\ge 1, \quad \langle \Delta_0^*\rangle=1,
\ee
where the sojourn probabilities $P_k(0)$ are known for many lattices in arbitrary dimension. The calculation of $\langle S_n^*\rangle$ as given in Eq.~(\ref{Scalc}) can now be carried out for any prescribed form of $\rho(n)$. For large values of $n$, this procedure becomes cumbersome and can be bypassed by the direct use of the generating function $S^*(\xi)$. Indeed, the behavior of this quantity in the limit $\xi\to 1^-$ yields the large $n$ behavior of $\langle S_n^*\rangle$ via a discrete Tauberian theorem (see p. 118 in Ref. \refcite{Hughes}).

The remainder of the present section deals with additional quantities of interest related to the arrival statistics of a mortal walker which can be straightforwardly computed in the framework of the generating function formalism.

\subsubsection{Expected number of sites visited at least a given number of times}
\label{sec:Sn}

Let us define $\langle S_n^{*(r)} \rangle$ as the average number of sites visited at least $r$ times in an $n$-step evanescent walk. Obviously, it is not possible to visit any given site $r\ge2$ times if no steps have been taken. We may therefore write $\langle S_0^{*(r)} \rangle=\delta_{1r}$.

Following the lines of the theory for immortal walkers we arrive at the relation
\be
\label{SrFr}
\langle S_n^{*(r)} \rangle = \sum_{j=1}^n F_j^{*(r-1)}(s_0|s_0) +
\sum_{s\neq s_0} \sum_{j=1}^n F_j^{*(r)}(s|s_0),
\ee
where $F_n^{*(r)}(s|s_0)$ is the probability that a mortal walker visits site $s$ for the $r$-th time when performing her $n$-th step given that she started the walk at $s_0$.  These probabilities are related to each other via the equation
\be
 F_n^{*(r)}(s|s_0)  = \sum_{j=1}^n   F_{j,n-j}^{*(r-1)}(s|s_0)  F_j^{*(r)}(s|s) ,\quad r\ge 2.
\ee
We hereafter adopt the convention followed by Hughes\footnote{
 This convention differs from the one followed by Montroll and Weiss \cite{Montroll1965} in that in the latter the start of the walk at $s_0$ is counted as the first visit to site $s_0$.} [see text below and formula (6.229) in Ref. \refcite{Hughes}], according to which \emph{the start of the walk is regarded as the zeroth visit to site}  $s_0$.  The generating function of $\langle S_n^{*(r)} \rangle $ is defined in the usual way:
\be
 S^{*(r)}(\xi)  = \sum_{n=0}^\infty \langle S_n^{*(r)} \rangle \xi^n= \delta_{1r}+\sum_{n=1}^\infty \langle S_n^{*(r)} \rangle \xi^n.
\ee
Hence, for $r>1$,
\be
\label{Srgt1}
 S^{*(r)}(\xi)  =  \sum_{n=1}^\infty \langle S_n^{*(r)} \rangle \xi^n, \quad r>1.
\ee
Introducing the corresponding generating function
\be
 F^{*(r)}(s|s_0;\xi)  = \sum_{j=1}^\infty   F_n^{*(r)}(s|s_0)   \xi^j,
\ee
one easily gets from \eqref{SrFr} and \eqref{Srgt1} the following relation,
\begin{align}
 S^{*(r)}(\xi)=\frac{1}{1-\xi}  \left\{ F ^{*(r-1)}(s_0|s_0;\xi)  +
 \sum_{s\neq s_0}   F^{*(r)}(s|s_0;\xi)   \right\}.
 \label{Sr*}
\end{align}

Knowledge of this quantity also allows one to infer the behavior of the number of sites $V_n^{*(r)} $ visited \emph{exactly} a given number of times $r$ in an $n$-step evanescent walk. Clearly,
\be
\langle V_n^{*(r)} \rangle= \langle S_n^{*(r)} \rangle - \langle S_n^{*(r+1)} \rangle,
\ee
implying
\be
V^{*(r)}(\xi)  = \sum_{n=0}^\infty \langle V_n^{*(r)} \rangle \xi^n = S^{*(r)}(\xi)  -  S^{*(r+1)}(\xi).
\ee

\subsubsection{Repeated visits to a given lattice site}

Let $ \beta_n^{*(r)}(s|s_0)$ be the probability that site $s$ is visited \emph{exactly} $r$ times in the first $n$ steps of the walk. Since $\sum_{j=1}^n F_j^{*(r)}(s|s_0)$ is the probability that site $s$ has been visited  \emph{at least} $r$ times, one clearly has
\be
\beta_n^{*(r)}(s|s_0)=\sum_{j=1}^n \left[ F_j^{*(r)}(s|s_0)-  F_j^{*(r+1)}(s|s_0)\right],
\ee
and the corresponding generating function reads
\be
\beta^{*(r)}(s|s_0;\xi)=\sum_{n=1}^\infty \beta_n^{*(r)}(s|s_0) \xi^n=\frac{1}{1-\xi}   \left[ F^{*(r)}(s|s_0;\xi)-  F^{*(r+1)}(s|s_0;\xi)\right].
\label{betaxi}
\ee
An associated quantity is the mean number $\langle \mu_n^*(s) \rangle$ of visits to site $s$ during the first $n$ steps of the walk,
\be
\langle \mu_n^*(s) \rangle=\sum_{r=1}^\infty r\, \beta_n^{*(r)}(s|s_0),
\ee
and its generating function
\begin{align}
\mu^*(s;\xi)&=\sum_{n=1}^\infty \langle \mu_n^*(s) \rangle \xi^n =\sum_{n=1}^\infty \xi^n \sum_{r=1}^\infty r\, \beta_n^{*(r)}(s|s_0) \nonumber \\
&=\sum_{r=1}^\infty r \sum_{n=1}^\infty \xi^n \beta_n^{*(r)}(s|s_0) =
\sum_{r=1}^\infty r   \beta^{*(r)}(s|s_0;\xi).
\end{align}

\section{Results for specific decay laws}
\label{Sec:Res}

\subsection{Exponential evanescence}

We now proceed to implement our results for the particular case of exponentially evanescent walkers, a case that allows explicit analytic calculation via the standard generating function formalism. Exponential evanescence is described by the exponentially decaying density $\rho(n) \propto \exp(-\lambda n)$, where $\lambda$ is the decay rate constant.  It is immediately evident that only for this step number dependence is the ratio $\rho(n)/\rho(m)$ a function of the difference $n-m$,  $\rho(n)/\rho(m) = \rho(n-m) \propto \exp[-\lambda(n-m)]$.  From Eq.~(\ref{P*Pgb}) it then follows that $P^*$ does not depend on the starting step $m$, that is, $P^*_{m,n-m}(s|s')=P^*_{n-m}(s|s')$. This greatly facilitates the subsequent calculations. We note also that the model is equivalent to considering an immortal walker stepping on a lattice where \emph{each} site is an imperfect trap whose (constant) absorption probability is $1-e^{-\lambda}$. A related model for a finite lattice has received some attention in Sec. 5 of Ref. \refcite{WalshKozak}, where the length of a walk of a random walker moving on an $N$-site lattice with $N-1$ imperfect traps and a \emph{single} perfectly absorbing trap was calculated.

\subsubsection{Number of distint sites visited}

We start with Eq.~(\ref{SD2}) together with Eq.~(\ref{D*F}),
\be
\label{SandF}
S^*(\xi) = \frac{1}{1-\xi} \left[ 1+\sum_{s\neq s_0} F^*(s|s_0;\xi)\right].
\ee
Next we focus on Eq.~(\ref{P*FP}), where we now take advantage of the independence of $P^*$ of the starting moment of the walk, so that
\be
\label{P*FPprime}
P_n^*(s|s_0)= \sum_{j=1}^n F_j^*(s|s_0) P_{n-j}^*(s|s),\quad n\ge 1.
\ee
This relation is now identical in form to that of a walk on a perfect lattice with non-evanescent walkers.  This means that we can follow the standard steps such as presented in chapter 3 of Ref. \refcite{Hughes} to ``turn the relation around" and arrive at
\be
\label{FPxi}
F^*(s|s_0;\xi)= \frac{P^*(s|s_0;\xi)-\delta_{ss_0}}{P^*(s_0|s_0;\xi)}.
\ee
This expression in Eq.~(\ref{SandF}) and  recognition of the easily proved consequence of normalization,
\be
\label{gendensity}
  \sum_{s}  P^*(s|s_0;\xi)=
  \sum_{n=0}^\infty  \rho(n)  \, \xi^n   \equiv  \rho(\xi),
\ee
leads to the result
\be
\label{SD5}
S^*(\xi)= \frac{1}{1-\xi} \frac{\rho(\xi)}{P^*(s_0|s_0;\xi)}.
\ee
In Eq.~(\ref{gendensity}) we have introduced the generating function $\rho(\xi)$ for the decay function
$\rho(n)$.

We can finally simplify our expression further by noting that for exponential evanescence the generating function is just
 \be
\rho(\xi)=\sum_{n=0}^\infty e^{-\lambda n} \xi^n = \sum_{n=0}^\infty \hat \xi^n=\frac{1}{1-\hat \xi},
\ee
where $\hat \xi = e^{-\lambda}\xi$. Furthermore, for exponential evanescence we also have a simple relation between the generating functions for return to the starting site with and without evanescence,
\begin{eqnarray}
\label{P*P}
 P^*(s|s_0;\xi)&=&\sum_{n=0}^\infty P_n^*(s|s_0)\xi^n
 =    \sum_{n=0}^\infty P_n(s|s_0) \rho(n)  \xi^n\nonumber\\
    & =& P(s|s_0;e^{-\lambda}  \xi)\equiv P(s|s_0;\hat \xi).
\end{eqnarray}
With this notation we can finally write
\be
\label{SD5ex}
S^*(\xi)= \frac{1}{1-\xi}\; \frac{1}{1-\hat \xi} \;\frac{1}{P(0;\hat \xi)},
\ee
and all we need to know in order to calculate this generating function is the generating function for returning to the site of origin in the absence of evanescence. We have explicitly recognized independence of the specific starting site by writing
\be
P^*(s_0|s_0;\xi)\equiv P^*(0;\xi)=P(0;\hat \xi).
\ee

\begin{figure}
\begin{center}
\includegraphics[width=0.6\textwidth]{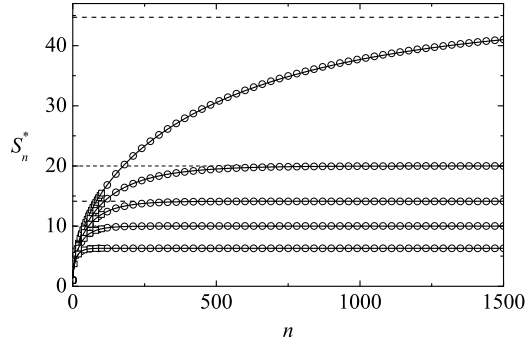}
\caption{\label{fig1} $\langle S_n^* \rangle$ vs. $n$ for a 1d lattice and, from top to bottom, 
$\lambda=0.001, 0.005, 0.01, 0.02$ and $0.05$. Solid lines, Eq. \eqref{SnLargen}; broken lines
give the values obtained from Eq. \eqref{Sinf1D}; circles, simulation values; squares, exact values obtained by identifying the first 100 coefficients in the $\xi$ power expansion of $S^*(\xi)$.}
\end{center}
\end{figure}

Equation~(\ref{SD5ex}) is the highlight of this subsection.
Lattice Green functions $P(0;\xi)$ are well known for the most common $d$-dimensional lattices (see for instance Refs. \refcite{Hughes} and \refcite{Guttmann}). The corresponding Green function in Eq.~(\ref{SD5ex}) allows one to obtain a large number of results for $S^*_n$. For example, identifying the $n$-th coefficient of the power series expansion of $S^*(\xi)$ one gets $S^*_n$. Besides, from Eq.~(\ref{SD5ex}) one immediately realizes that, in contrast to the classical case of immortal walkers, the average number of sites visited by a mortal P\'olya walker with exponential evanescence as $n$ goes to infinity is a \emph{finite} quantity given by
\be
\label{Sinfgral}
 \langle S_\infty^* \rangle =  \frac{1}{1-e^{-\lambda}} \;\frac{1}{P(0;e^{-\lambda})}.
\ee
For an infinite one-dimensional lattice one has $P(0;\xi)=(1-\xi^2)^{-1/2}$. Employing this in
Eq.~(\ref{Sinfgral}) yields 
\be
\label{Sinf1D}
\langle S_\infty^* \rangle= [(1+e^{-\lambda})/(1-e^{-\lambda})]^{1/2}.
\ee
The large $n$ behavior can be shown to be given by the formula 
\be
\label{SnLargen}
\langle S_n^* \rangle \approx \langle S_\infty^* \rangle - \sqrt{2}\,
\frac{ I_{e^{-\lambda}}(n+1,1/2)}{(1-e^{-\lambda})^{1/2}}
\ee
where $I_x(a,b)$ stands for the regularized beta function. As seen in Fig. \ref{fig1}, this formula is in excellent agreement with numerical simulations.

For two-dimensional regular lattices the generic form of the generating function is $P(0;\xi)\sim A/\pi \ln[B/(1-\xi)]$ for $\xi\to 1^-$ where $A$ and $B$ are constants which depend on the details of the lattice geometry.\cite{Hughes} From Eq.~(\ref{Sinfgral}) we find
 \be
\label{S2d}
 \langle S_\infty^* \rangle \sim   \frac{\pi}{\lambda\, A \log(B/\lambda)}, \quad \lambda\to 0.
\ee
In particular, one has e.g. $A=1$ and $B=8$ for a square lattice and $A=\sqrt{3}/2$ and $B=12$ for a triangular lattice.

It is well-known that the probability $\mathcal{R}$ of return to the origin for an immortal walker in $d\ge 3$ is intimately related to the generating function $P(0;\xi)$ evaluated at $\xi=1$, i.e.,  $\mathcal{R}=1-1/P(0;1)$. Eq.~(\ref{Sinfgral}) then yields $ S_\infty^* \sim (1-\mathcal{R}) \lambda^{-1}$ as $\lambda\to 0$. It is interesting to note that for $d\ge 2$ the value of  $\langle  S_\infty^* \rangle $ as $\lambda\to 0$ can also be obtained from the main asymptotic term of $ \langle S_n \rangle $  by replacing the number of steps $n$ by the average number of steps taken by the walker before it disappears, i.e, by $1/\lambda$.  This hand-waving approximation does not work for $d=1$ as, in this case, $\langle S_\infty^* \rangle \sim \sqrt{2/\lambda}$ whereas $\langle S_{1/\lambda}\rangle \sim \sqrt{8/(\pi\lambda)}$.

\subsubsection{Repeated visits to a given lattice site}

\begin{figure}
\begin{center}
\includegraphics[width=0.6\textwidth]{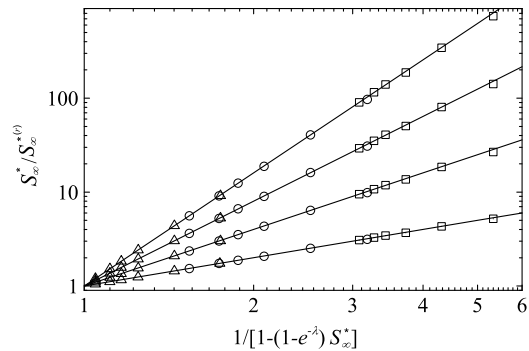}
\caption{\label{fig:SnrSn1}
$\langle S_\infty^{*(r)}\rangle$ vs  $\langle S_\infty^{*} \rangle$ and $\lambda$ for several values of $r$ and  $\lambda$ and three different lattices. Symbols: numerical simulations for $d=1$ (triangles), $d=2$ (square lattice, circles), and  $d=3$ (cubic lattice, squares) for $10^5$ runs. The values of $\langle S_\infty^{*(r)} \rangle$ are obtained from the simulation of $\langle S_n^{*(r)} \rangle $ with $n$ sufficiently large to observe no change in at least three significant figures. From left to right:  $\lambda=0.1, 0.05, 0.03, 0.01, 0.05, 0.001$, with  from top to bottom $r=2, 3, 4, 5$. The straight lines of slope $(r-1)$ through the origin are the theoretical predictions. }
\end{center}
\end{figure}

Taking Eq.~\eqref{Sr*} as a starting point and following the procedure of Montroll and Weiss \cite{Montroll1965}, one easily finds that the generating function for the average number of sites visited at least $r$ times,  $\langle S_n^{*(r)} \rangle$, is
 \be
\label{Sr2}
 S^{*(r)}(\xi)=
  \left[1- \frac{1}{P(0; \hat{\xi)}}\right]^{r-1} S^{*}(\xi).
\ee
From here one  finds formulas \cite{Montroll1965} for $\langle S_n^{*(r)} \rangle $ in terms of $\langle S_n^{*} \rangle $ . For instance, in dimension $d=1$ one obtains  $\langle S_n^{*(2)} \rangle =  \langle S_n^{*} \rangle-1-e^{-\lambda}$,
$\langle S_n^{*(3)} \rangle = 2 \langle S_n^{* } \rangle -e^{-2\lambda}  \langle S_n^{* } \rangle - 2 -2e^{-\lambda}$, etc.
For the average number of revisits to the origin after $n$ steps, $\langle \mu_n^* (s_0) \rangle$, one finds the generating function
$\mu^*(s_0;\xi)=  (1-\xi)^{-1}  [P(0;\hat{\xi}) -1]$.  (We follow the convention of \emph{not} counting the initial occupancy of the origin by the walker at step zero as the first revisitation \cite{Hughes}). We find that the asymptotic average number of revisits to the origin in any dimension is given by $\langle \mu_\infty^*(s_0)\rangle=[(1-e^{-\lambda}) \langle S_\infty^* \rangle]^{-1}-1$, while the average number of visits to a site $s$ other than the origin is given by $\langle \mu_\infty^*(s) \rangle =P(s|s_0;e^{-\lambda})$, one of the few previously known results for exponentially evanescent walkers (see subsection 3.2.4 in Ref.~\refcite{Hughes}).
The average number of sites visited $r$ times before the walker dies is in any dimension given  by
\be
\langle S_\infty^{*(r)} \rangle=[(1-e^{-\lambda}) \mu_\infty^*]^{r-1} \, (\langle S_\infty^* \rangle)^r.
\label{SrS3}
\ee
This result has been tested with numerical simulations for different values of $r$ and of $\lambda$, and extremely good agreement between simulations and theory has been found (cf. Fig. \ref{fig:SnrSn1}).

\subsection{Power-law evanescence}
\label{sub:Plev}

In what follows we consider power-law evanescence, i.e., a decay law of the form $\rho(n)=(1+\lambda n)^{-\beta}$ with $\lambda>0$ and  $\beta>0$. We directly use $\langle S_n^*\rangle=\sum_{m=0}^n \rho(n) \langle \Delta_n \rangle$ and rely on our knowledge of  $\langle \Delta_n \rangle$ for large and small $n$ for the most common lattices.\cite{Hughes,Montroll1965} For example, $\langle \Delta_n \rangle \sim
  (1-\mathcal{R})\, \lambda^{-\beta} n^{-\beta} \left(1+C\, n^{-1/2} +\cdots\right)$ for three-dimensional lattices and large $n$, where $C$ is a constant. Because $\rho(n)\sim (\lambda n)^{-\beta}$ for large $n$, one sees immediately that $\langle S_\infty^* \rangle$ is finite for $\beta>1$.  For the case of slow evanescence  ($\lambda\to 0$)  it is not difficult to find that this finite value is $\langle S_\infty^*\rangle \sim (1-\mathcal{R})/[(\beta-1) \lambda]$. However, for $\beta<1$  the result for slow evanescence is quite different:
\be
\label{Sn3dpot}
\langle S_n^* \rangle \sim \frac{1-\mathcal{R}}{1-\beta} \, \lambda^{-\beta}\,n^{1-\beta}
\ee
for large $n$. For $\beta=0$ (no evanescence) one recovers the classical result \cite{Hughes}. For the marginal case $\beta=1$ one obtains
\be
\label{Sn3dmarg}
\langle S_n^* \rangle \sim (1-\mathcal{R}) \lambda^{-1 }\log n.
 \ee
This way we discover that the average number of distinct sites visited by a mortal walker before it dies is infinite for $\beta\ge 1$, whereas  this quantity is finite for $\beta<1$.  This is also true for two-dimensional lattices, and for $d$-dimensional lattices with $d\ge 4$.  However, for the one-dimensional lattice the  critical value is  $\beta=1/2$. 

\section{The target problem with mortal traps}
\label{Sec:tp}

\subsection{Discrete-time problem}

The quantities we have presented so far deal with mortal walkers in a perfect lattice, but there are many situations in which the lattice includes an imperfection of some sort.  In particular, one can consider a collection of initially randomly distributed, independently moving walkers (``the traps'') distributed on a lattice with a target placed at a given site (the imperfection). The target is instantaneously killed as soon as any of the traps steps on the target site, and one may wish to know the survival probability $Q_T^*(n)$ of the target up to a given step $n$. This is the discrete version of  the so- called target problem, but it differs from the typical setting in that our traps are now mortal. The target problem is clearly a first-passage problem, as $Q_T^*(n)$ is the probability that none of the traps has visited the target site up to time step $n$. Since we are not interested in the fate of the trap that kills the target after it has ``accomplished its goal'', for practical purposes we can consider the target site as an absorbing site that terminates the trap's trajectory.

Now, it is well known that in the case of immortal traps the survival probability of the target, $Q_T(n)$, is related to $\langle S_n \rangle$,  the average number of distinct sites visited by a \emph{single} trap in the corresponding perfect lattice, i.e., a lattice where the absorbing site is replaced by a regular site. One then has the equality
\be
Q_T(n)=\exp\left[- c_0 (\langle S_n \rangle-1) \right].
\label{QTnSn}
\ee
Here $c_0$ is the (constant) number density of traps, that is, the fraction of lattice sites initially occupied by traps. This relation is very useful, since knowledge of $\langle S_n \rangle$, a quantity calculated for a perfect lattice,  leads to the determination of $Q_T(n)$ for a defect lattice and the corresponding reaction rate.\cite{AYLJMMNP} In this case, the target eventually dies with certainty, since $\langle S_n \rangle \to \infty$ as $n\to \infty$ in all dimensions. Since the link between $Q_T(n)$ and $\langle S_n \rangle$ is purely geometric, one can thus conjecture that a similar relation also holds in the case of mortal traps,
\be
Q^*_T(n)=\exp\left[- c_0 (\langle S^*_n \rangle-1) \right],
\ee
where $c_0$ is the \emph{initial value} of the number density of traps; this number now decreases in time. This relation can indeed be rigorously proven. According to the results displayed in the previous section, for exponential evanescence $Q^*_T(n)$ goes to a finite value given by Eq.\eqref{Sinfgral} as $n\to\infty$. Hence, the target has a non-zero chance of survival in all dimensions, as opposed to the classic case with immortal traps. In contrast, when one has power-law evanescence, the target may or may not survive depending on the value of the decay exponent $\beta$ (see subsection \ref{sub:Plev}).

\subsection{Continuous space and time}
\label{Subsec:CTP}

The target problem can also be formulated in continuous rather than discrete space and time.  For this purpose, it is convenient to consider the relation between $\langle S_n\rangle$ and the mean volume explored by an immortal trap in continuous space up to a given time, $\langle v\rangle_t$ (we are here adopting  standard notation). The latter is defined in terms of the Wiener sausage generated by a spherical walker of radius $R$ in time $t$. When there is no evanescence, this relation is well known and in the long time limit it reads as follows:\cite{Berezhkovskii1989}
\be
\langle v\rangle_t \sim B \ell^d\langle S_n\rangle,
\label{vtSn}
\ee
where $\ell$ is the lattice constant in the discrete lattice and $d$ is the dimensionality.  The constant prefactor $B$ depends on $R$, $\ell$, dimensionality and the specific lattice structure. The relation is based on the correspondence $n=2dDt/\ell^2$ when $n\gg 1$ and $R\gg \ell$ (see appendix \ref{appTD}).
For mortal walkers, the relation between the number of distinct sites visited and the explored volume does not change, i.e., $\langle S_n\rangle$ and $\langle v\rangle_t$ are respectively replaced by $\langle S_n^*\rangle$ and $\langle v^*\rangle_t$ in Eq. \eqref{vtSn}. Explicit results in continuous space for the survival probability $Q^*_T(t)$ can be found in Ref. \refcite{AYLJMMNP} for both diffusive and subdiffusive walkers.

We have recently suggested \cite{YALPRL13} that trap evanescence may provide an alternative explanation for the behavior of certain defect-mediated relaxation processes, as in the celebrated model of dielectric relaxation first introduced by Glarum \cite{Glarum} in one dimension and later refined by Bordewijk to account for many-body effects both in one and three dimensions.\cite{Bordewijk} In such a setting, the relaxation function of a given configuration of a target molecule displays stretched exponential behavior (so-called Kohlrausch-Williams-Watts behavior) when the molecule is subject to a flux of diffusing defects. The identification of these defects (i.e., carriers of free volume or occurrences of elementary relaxation events) as \emph{the traps} is the basis of the defect diffusion model to explain the observed behavior of the survival probability $Q_T(t) \sim \exp(-c_0 \langle v\rangle_t)$ of the molecular configuration, where $c_0$ is the defect concentration. The observed experimental behavior is $\ln Q_T(t) \sim t^\theta$, where $\theta$ may take a wide range of values. However,  only the values $\theta=1/2$ or $\theta=1$ are possible for normal diffusive defects because  $\langle S_n \rangle \propto \langle v\rangle_t\propto t^{1/2}$ for $d=1$ and (and then $\theta=1/2$) and $\langle S_n \rangle \propto \langle  v\rangle_t\propto t$ for $d\ge 2$ (and then $\theta=1$). The Glarum-Bordewijk model was extended by  Shlesinger and Montroll \cite{ShlesingerMontroll}   by assuming that the movement of the defects may be described by a CTRW model with a power-law waiting time distribution  $\sim t^{-1-\gamma}$, $0<\gamma<1$, which leads to $\theta=\gamma/2$ for $d=1$ and $\theta=\gamma$ for $d\neq 1$ \cite{Scher1991}.  Thus, the stretched exponential relaxation with $\theta\neq 1/2$ is explained by assuming the diffusion of the defects to be anomalous with anomalous diffusion exponent $\gamma$.

Our results suggest that it may not be necessary to invoke anomalous diffusion to explain stretched exponential behavior, as evanescence of defects may lead to a similar qualitative behavior even when defect diffusion is normal. Such evanescence events have indeed been observed in experiments \cite{Heggen,Chakravarty2009}. As we have shown, different evanescent behaviors lead to different kinds of relaxation; for example,  from Eq.~\eqref{Sn3dpot} we see that $\langle v^*\rangle_t \propto t^{1-\beta}$ for $\beta<1$ for $d\ge 3$, so that one can get stretched exponential relaxation with exponent $\theta=1-\beta$ when the concentration of defects decays as a power law. Moreover, when $\beta=1$, implying that the concentration $c$ of defects decays as $\rho(t)\sim 1/t$ for large $t$ (a decay found in some bimolecular reactions in condensed media), one gets from \eqref{Sn3dmarg} that $\langle v^*\rangle_t \propto \ln t$, which in turn leads to \emph{algebraic} relaxation.\cite{BluKlaZuOptical, ShlesingerJCP79, ShlesingerJSP84}. This leads one to hypothesize that a situation where the interaction with the target is delayed because of the disappearance of the traps might become experimentally indistinguishable from a scenario where such a delay is caused by the subdiffusive motion of the traps.

Finally, it is worth recalling that the survival probability of the target in the target problem can be considered as a first approximation (Rosenstock approximation) to the survival probability of the target in the so-called trapping problem, in which the target diffuses and the mortal walkers become immobile traps. Our results can also be applied to this fundamental problem when the concentration of traps decreases with time. For the continuous time version of the trapping problem, Den Hollander and Shuler calculated the survival probability of the target in the regime of validity of the Rosenstock approximation. \cite{HollanderShuler92} It is worth pointing out that their result for the survival probability is a consequence of the behavior of $\langle v^*\rangle_t$ in the long time regime.

\section{Conclusions and Outlook}
\label{Sec:Conc}

We have reviewed some recent results for the first arrival statistics of mortal walkers, a problem which has received very little treatment in the literature. \emph{A posteriori}, the approach used seems straightforward, yet the underlying physics is very different from the case of immortal walkers, thus making this problem a very interesting one. Remarkably enough, the profound modifications introduced by the evanescence reaction in the behavior of  evanescent walkers can be studied using the generating function formalism originally developed for immortal walkers. As we have seen, the evanescence reaction dramatically changes the behavior of a number of characteristic quantities, e.g., the distinct number of sites visited as a function of time, or the survival probability of a target surrounded by randomly moving traps. For instance, the average number of distinct sites visited by an immortal walker goes to infinity as $n\to\infty$, while (depending on the speed of evanescence) it may be finite for an evanescent walker. Consequently, the survival probability of an immobile target in the presence of mobile traps is zero if the traps are immortal; however, if the traps evanesce at a sufficiently rapid rate, the target has a chance of survival.

Our results concerning the target problem also provide an alternative explanation to the stretched exponential relaxation observed in the defect diffusion model. In previous work, the possibility of anomalous diffusion induced by long-tailed waiting time distributions was invoked to explain the observed behavior, but we have seen that the same behavior is obtained with normally diffusive defects provided that these may disappear at a sufficiently slow rate. Indeed, there is evidence for a decrease in the concentration of defects in time due to different processes such as, for instance, defect coalescence or annihilation at sinks. However, since subdiffusion and evanescence do not exclude each other, both of them could play a role in conjunction and should therefore both be taken into account for quantitative studies.

Finally, the aspects studied here by no means exhaust the family of problems related to explored territory (or distinct sites visited) by mortal walkers. Studies for the case of biased walks, continuous-time random walks, walks in confined spaces, and L\'{e}vy flights and walks are underway and will be presented elsewhere.

\section*{Acknowledgments}

This work was partially funded by the Ministerio de Ciencia y Tecnolog\'ia (Spain) through Grant No. FIS2010-16587 (partially financed by FEDER funds), by the Junta de Extremadura through Grant. No. GRU10158, and by the US National Science Foundation under Grant No. PHY-0855471.

\begin{appendix}[Derivation of a continuum reaction-diffusion equation from the master equation for mortal walkers]
\label{appRDE}

We start from the fundamental equation \eqref{evanescentME} and introduce the probability density function or ``concentration'' $c$ via the definition $P_n^*(s|s_0)=\Delta x \, c(s\Delta x, n\Delta t| s_0\Delta x)$,
where $\Delta x$ is the lattice constant. Employing this definition in \eqref{evanescentME} we get
\be
\label{smallpeq}
c(x, t+\Delta t| x_0)=\frac{1}{2}\left[c(x-\Delta x, n\Delta t| x_0)
+c(x+\Delta x, t| x_0)\right]\frac{\rho(t+\Delta t)}{\rho(t)},
\ee
where the notation $x=s\Delta x, x_0=s_0\Delta x$ and $t=n\Delta t$ has been used.
We now perform the usual Taylor expansion in terms of $\Delta x$ and $\Delta t$ as follows:
\begin{subequations}
\begin{align}
c(x, t+\Delta t| x_0)& \approx c(x, t| x_0)+\Delta t \,\partial_t c(x,t), \\
c(x+\Delta x, t| x_0)& \approx c(x, t| x_0)+\Delta x \,\partial_t p(x,t)+\frac{(\Delta x)^2}{2}\,\partial_x^2 c(x,t),\\
\rho(t+\Delta t)& \approx \rho(t)+\Delta t \,\dot{\rho}(t).
\end{align}
\end{subequations}
Inserting these expressions into \eqref{smallpeq}, dividing the resulting equation by $\Delta t$ and letting
$\Delta t$ and $\Delta x$ simultaneously go to zero while keeping the ratio $(\Delta x)^2/\Delta t$ fixed, we get the following continuum reaction-diffusion equation:
\be
\label{contRDE}
\partial_t c(x,t|x_0)=D\,\partial_x^2 c(x,t|x_0)-\hat{\lambda}(t)\,c(x,t|x_0),
\ee
where the diffusion coefficient is taken to be $D\equiv \lim_{\Delta x,\Delta t\to 0}(\Delta x)^2/2\Delta t$ and the negative logarithmic derivative $\hat{\lambda}(t)=-\dot{\rho}/\rho$ is a time-dependent rate constant. The latter is assumed to be small in absolute value so that it makes sense to take the diffusion limit (this limit implies that, on average, a sufficient number of time steps $n \gg 1$ must have been taken before reaction takes place). The particular case of exponential evanescence $\rho(t)=\rho_0\,\exp{(-\hat{\lambda} \, t)}$ leads to a time-independent rate constant $\hat{\lambda}$.  The above derivation can be straightforwardly generalized for non-nearest neighbor walks or walks in higher dimensions.
\end{appendix}

\begin{appendix}[Time-discretization of the walker density]
\label{appTD}

The mortal walker performs steps on the $d$-dimensional lattice at regular time intervals $\Delta t$, hence the stepping times are $t_n=n\Delta t$. In the case of an exponential evanescence process, its density decay is given by
\be
\rho(t)=e^{-\hat \lambda t } \to \rho(t_n) \equiv \rho(n)=\rho(-\hat \lambda t_n),
\ee
i.e.,
\be
\rho(n)=\rho(0)\, e^{-\lambda n},
\ee
where we have introduced $\lambda \equiv \hat \lambda  \Delta t$.

One can proceed in a similar way with a power law decay. One has
\be
\rho(t)=\frac{\rho_0}{(1+\hat \lambda  t)^\beta}  \to
\rho(n)=  \frac{1}{(1+\lambda n)^\beta}
\ee
\be
\rho(n)=\frac{1}{(1+n  \hat \lambda \Delta t )^\beta} \Rightarrow \lambda= \hat \lambda \Delta t
\ee
Alternatively, the relation between $\lambda$ and $\hat \lambda$ can be expressed in terms of the diffusion coefficient
\be
D=\frac{1}{2d} \frac{\ell^2}{\Delta t},
\ee
where $\ell^2$ is the variance of the step length distribution. Using the resulting expression for $\Delta t$, one gets
\be
\lambda= \hat \lambda \Delta t = \hat \lambda   \, \frac{\ell^2  }{2d D}.
\ee
Conversely, to obtain the continuum limit one must perform the following replacements
\be
t_n\to t \mbox{\qquad and \qquad} n= \frac{t_n}{\Delta t} \to t \, \frac{2d D}{\ell^2}.
\ee
We make use of this correspondence in \ref{Subsec:CTP}.
\end{appendix}

\end{document}